\documentclass[a4paper,11pt]{article}
\usepackage{pos}
\usepackage{graphicx}
\usepackage{tabularx}
\usepackage{booktabs}
\usepackage{amsmath}
\usepackage{multicol}
\usepackage{tikz}
\usepackage{setspace}
\hypersetup{colorlinks,citecolor=blue,linkcolor=blue,urlcolor=blue}

\newcommand{\orcidPS}{\href{https://orcid.org/0000-0002-5976-0317}{\includegraphics[height=9pt]{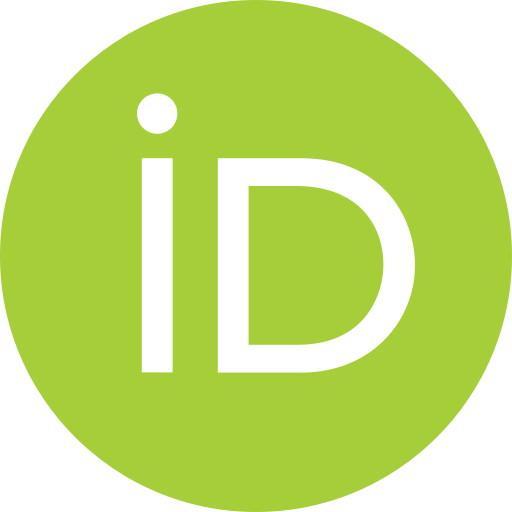}~}}
\newcommand{\orcidJR}{\href{https://orcid.org/0000-0003-1866-0157}{\includegraphics[height=9pt]{orcid.png}~}}
\newcommand{\orcidPB}{\href{https://orcid.org/0000-0003-4579-0387}{\includegraphics[height=9pt]{orcid.png}~}}

\title{Automated NLO SM corrections for all colliders}

\author*[a]{{Pia Bredt}\orcidPB}
\author[a]{J\"urgen Reuter\orcidJR}
\author[a]{Pascal Stienemeier\orcidPS}
\affiliation[a]{Deutsches Elektronen-Synchrotron DESY, \\
	Notkestr. 85,
	22607 Hamburg, Germany}

\emailAdd{pia.bredt@desy.de}
\emailAdd{juergen.reuter@desy.de}
\emailAdd{pascal.stienemeier@desy.de}

\abstract{We summarize the status of automated NLO SM corrections for hadron and lepton collider processes in the multi-purpose event generator WHIZARD. The focus will be on NLO EW and QCD-EW mixed corrections at the LHC. Also, recent progress on the inclusion of EW corrections in future lepton collider processes and on the POWHEG-matched event generation in the NLO automated setup will be discussed.\\
	
	DESY-22-151}

\FullConference{%
  41st International Conference on High Energy physics - ICHEP2022\\
  6-13 July, 2022\\
  Bologna, Italy
}

\begin{document}
\maketitle

\section{Introduction -- Automated NLO corrections in WHIZARD}
Monte-Carlo event generators in general are indispensable tools for simulating physics in a universal manner with a predictive power at the level of exclusive data. \texttt{WHIZARD} \cite{Kilian:2007gr} is a multi-purpose  generator for cross sections, distributions and simulated event samples for lepton and hadron collider processes covering SM and BSM physics. It provides an intrinsic tree-level matrix element generator \texttt{{O'Mega}} \cite{Moretti:2001zz} and a phase-space integrator with MPI-based parallelization capabilities \texttt{{VAMP2}} \cite{Brass:2018xbv}. This automated framework recently has been extended to the stage of accounting for complete perturbative NLO corrections in the full SM, i.~e. NLO QCD, EW and mixed corrections. The regularization of infrared singularities is based on the FKS subtraction scheme where one-loop virtual matrix elements are accessed by generic interfaces to one-loop providers such as \texttt{OpenLoops} \cite{Buccioni:2019sur}, \texttt{RECOLA} \cite{Actis:2012qn} and \texttt{GoSam} \cite{Cullen:2014yla}.
The matching of NLO QCD event generation to parton showers in \texttt{WHIZARD} happens via the POWHEG matching scheme.

We defer a detailed documentation of the complete NLO module, involving the extension of the NLO QCD automated framework \cite{Bach:2017ggt,ChokoufeNejad:2016qux}, to a separate publication. We summarize recent developments towards the NLO EW automated setup as well as the POWHEG matching below.
\section{NLO EW and SM mixed corrections at the LHC}
Automated NLO EW corrections for LHC processes in general impose new computational challenges in theoretical predictions.
First of all, photon-induced processes as well as QED IR-safety criteria, e.~g. requiring photon recombination, play a role already for pure EW processes.
Concerning off-shell vector boson processes of this class the \texttt{WHIZARD+OpenLoops} NLO EW framework is validated with \texttt{MG5\_aMC@NLO} \cite{Frederix:2018nkq}. Results of these checks for a selection of benchmark processes including neutral- and charged-current processes (with and without associated Higgs), VBF as well as single top plus jet processes are listed in Table~\ref{offshellVB}. The input parameters for the checks presented in this section are chosen according to the setup in ref.~\cite{Frederix:2018nkq}, particularly,

{\quad$\sqrt{s}= 13$ TeV, \quad $\mu_R=\mu_F=\frac{H_T}{2}=\frac{1}{2}\sum_{i}\sqrt{p_{T,i}^2+m_i^2}$, \quad $\alpha$ input scheme: $G_{\mu}$ CMS}.
\begin{table}[h]
	\centering
	\scriptsize
	\begin{tabularx}{0.73\textwidth}{l|r|r|r|r|r}
		process  & $\alpha^n$ &\texttt{MG5\_aMC@NLO} $\sigma_{\text{NLO}}^{\text{tot}}$ [pb] &\texttt{WHIZARD} $\sigma_{\text{NLO}}^{\text{tot}}$ [pb] & $\delta$ [\%]  & $\sigma^{\text{sig}}_{\text{NLO}}$\\
		$pp \rightarrow$ &&&\texttt{+OpenLoops} \quad\quad\quad\quad&&\\
		\hline
		$e^+\nu_e\mu^-\bar\nu_\mu$  & $\alpha^4$ &     $ 0.52794(9) $   &   $ 0.52816(9) $ & $ +3.69 $  &  $ 1.69 $\\
		$e^+e^-\mu^+\mu^-$ & $\alpha^4$ &          $ 0.012083(3) $    &     $ 0.012078(3) $     & $ -5.25 $ & $ 1.26 $\\
		$H e^+\nu_e$   &     $\alpha^3$    &  $ 0.064740(17) $       &     $ 0.064763(6) $   &   $ -4.04 $ & $ 1.24 $  \\
		$ H e^+e^-$  & $\alpha^3$  &   $ 0.013699(2) $         &     $ 0.013699(1) $   &   $ -5.86 $  &$ 0.32 $ \\ 
		$ H jj$     &  $\alpha^3$      &     $ 2.7058(4) $     &      $ 2.7056(6) $   &  $- 4.23 $  & $ 0.27 $ \\
		$tj$     &   $\alpha^2$     &   $ 105.40(1) $      &    $ 105.38(1) $   &  $ -0.72 $ &  $ 0.74 $
	\end{tabularx}
\caption{Checks for a set of pure EW off-shell vector boson processes at the LHC with $\delta\equiv\sigma^{\text{tot}}_{\text{NLO}}/\sigma^{\text{tot}}_{\text{LO}}-1$}
\label{offshellVB}
\end{table}

Furthermore, for processes with on-shell bosons $VV$, $VH$, $VVV$, $VVH$ and $VHH$ with $V=W^{\pm},Z$ agreeing NLO EW cross section results with those of \texttt{MUNICH/MATRIX} \cite{Kallweit:2014xda,Buonocore:2021rxx} using \texttt{OpenLoops} could be achieved for relative MC errors well-below sub-per-mille level.

For processes at $\mathcal{O}(\alpha_s)$ and higher NLO fixed order contributions possibly contain overlapping QCD-EW corrections demanding for the subtraction of both, QED and QCD IR singularities. More precisely, both correction types of possible IR splittings have to be considered in the counterterms for all NLO contributions at fixed coupling powers except for those at orders leading in $\alpha_s$ or $\alpha$.
%
  	\begin{table}
	\centering
	\scriptsize
	\begin{tabularx}{0.77\textwidth}{l|r|r|r|r|r}
		\multicolumn{1}{c}{ }&& \multicolumn{2}{c|}{$\sigma^{\text{tot}}$ [fb]}   & $\sigma^{\text{sig}}$ & \textit{rel. deviation}\\
		$pp \rightarrow t\bar{t}H$ & $\alpha_s^m\alpha^n$& \multicolumn{1}{c|}{\texttt{MUNICH+OpenLoops}}
		&\texttt{WHIZARD+OpenLoops} & &\\
		\hline
		$\text{LO}_{21}$       & $\alpha_s^2\alpha$ & \multicolumn{1}{r|}{      $  3.44865(1)\cdot 10^2 $ }      &    $  3.4487(1)\cdot 10^2 $    &  $ 0.76$ & $ 0.003\%$ \\
		$\text{LO}_{12}$       & $\alpha_s\alpha^2$ &        \multicolumn{1}{r|}{ $  1.40208(2)\cdot 10^0 $ }      &  $  1.4022(1)\cdot 10^0 $  &     $ 1.44 $ & $0.011\%$ \\
		$\text{LO}_{03}$       & $\alpha^3$ & \multicolumn{1}{r|}{ $  2.42709(1)\cdot 10^0 $   }   &    $ 2.4274(2) \cdot 10^0 $    & $ 2.07 $ & $0.011\%$ \\
		\hline
		$\text{NLO}_{31}$       & $\alpha_s^3\alpha$   &      $  9.9656(4)\cdot 10^{1} $      &    $ 9.968(4) \cdot 10^1 $    &$ 0.62$ & $ 0.023\%$\\
		$\text{NLO}_{22}$       & $\alpha_s^2\alpha^2$   &      $ 6.209(1) \cdot 10^0 $     &    $  6.208(2)\cdot 10^0 $     &$ 0.20$ & $ 0.009 \%$\\
		$\text{NLO}_{13}$       & $\alpha_s\alpha^3$  &      $  1.7238(2)\cdot 10^0 $      &    $ 1.7232(5) \cdot 10^0 $     &$1.24 $ & $0.040\%$\\
		$\text{NLO}_{04}$       & $\alpha^4$  &      $  1.5053(3)\cdot 10^{-1} $     &    $  1.5060(7)\cdot 10^{-1}  $     &$1.00 $ & $0.048\%$
	\end{tabularx}
\caption{Comparison of all fixed order LO and NLO contributions to the cross section of $pp\to t\bar{t}H$}
\label{ppttH}
\end{table}
By appropriate modifications of the FKS automated framework in \texttt{WHIZARD} the contributions of the whole tower of coupling powers can be computed. By interfacing \texttt{OpenLoops} process libraries the validation of all $\alpha_s$ leading and subleading NLO contributions of $pp\rightarrow t\bar{t}(H/Z/W^{\pm})$ has been performed with reference results of \texttt{MUNICH/MATRIX}. Results for $pp\to t\bar{t}H$ are exemplarily shown in Table \ref{ppttH}.
Due to IR-safety criteria photons have to be treated on a democratic basis with gluons in the jet definition. This yields a non-trivial phase-space cut evaluation including photon recombination and jet clustering for processes with jets and leptons in the final state. In this context, checks with \texttt{MG5\_aMC@NLO} for $pp\rightarrow e^+\nu_e j$ and $pp\rightarrow e^+e^- j$ are performed with results listed in Table~\ref{ppllj}.

	\begin{table}[h]
		\centering
		\scriptsize
		\begin{tabularx}{0.8\textwidth}{l|r|r|r|r|r}
			$pp \rightarrow$  & $\alpha_s^m\alpha^n$ &\texttt{MG5\_aMC@NLO} $\sigma_{\text{NLO}}^{\text{tot}}$ [fb] &\texttt{WHIZARD+OpenLoops} $\sigma_{\text{NLO}}^{\text{tot}}$ [fb] & $\delta$ [\%]  & $\sigma^{\text{sig}}_{\text{NLO}}$\\
			\hline
			$e^+\nu_ej$          & $\alpha_s\alpha^2$ &      $ 9.0475(8)\cdot10^5 $        &       $ 9.0459(7)\cdot 10^5 $     & $ -1.11 $ &  $ 1.5 $\\
			$e^+e^-j$       & $\alpha_s\alpha^2$ &       $ 1.4909(2)\cdot 10^5 $      &    $ 1.4908(2)\cdot 10^5 $    & $ -1.00 $ & $ 0.4 $
		\end{tabularx}
	\caption{Checks for cross sections at NLO EW for representative processes requiring involved cuts}
	\label{ppllj}
	\end{table}
\section{Lepton collider processes at NLO EW}
	Under certain conditions predictions for lepton collision processes at NLO EW are reliable in a fixed order massive initial state approximation. Using NLO EW accurate amplitudes from \texttt{RECOLA} this is automated in \texttt{WHIZARD} with adjusted FKS phase space construction for massive initial-state emitters.
	This framework is used in our recent study on NLO EW corrections to multi-boson production at a future muon collider \cite{Bredt:2022dmm}.
For a universal treatment of collinear ISR effects in NLO calculations NLL accurate electron PDFs \cite{Frixione:2019lga,Bertone:2019hks} - implemented and validated in \texttt{WHIZARD} - have to be applied. Embedding these into the FKS framework is under development. Numerical pitfalls are posed by the interplay of FKS ISR construction and the divergent behavior of the PDFs in the asymptotic $z\to1$ limit which is being handled by adequate phase-space mappings.
\section{POWHEG-matched and showered NLO event generation}
\begin{figure}
  		\centering
	\includegraphics[width=0.3\textwidth]{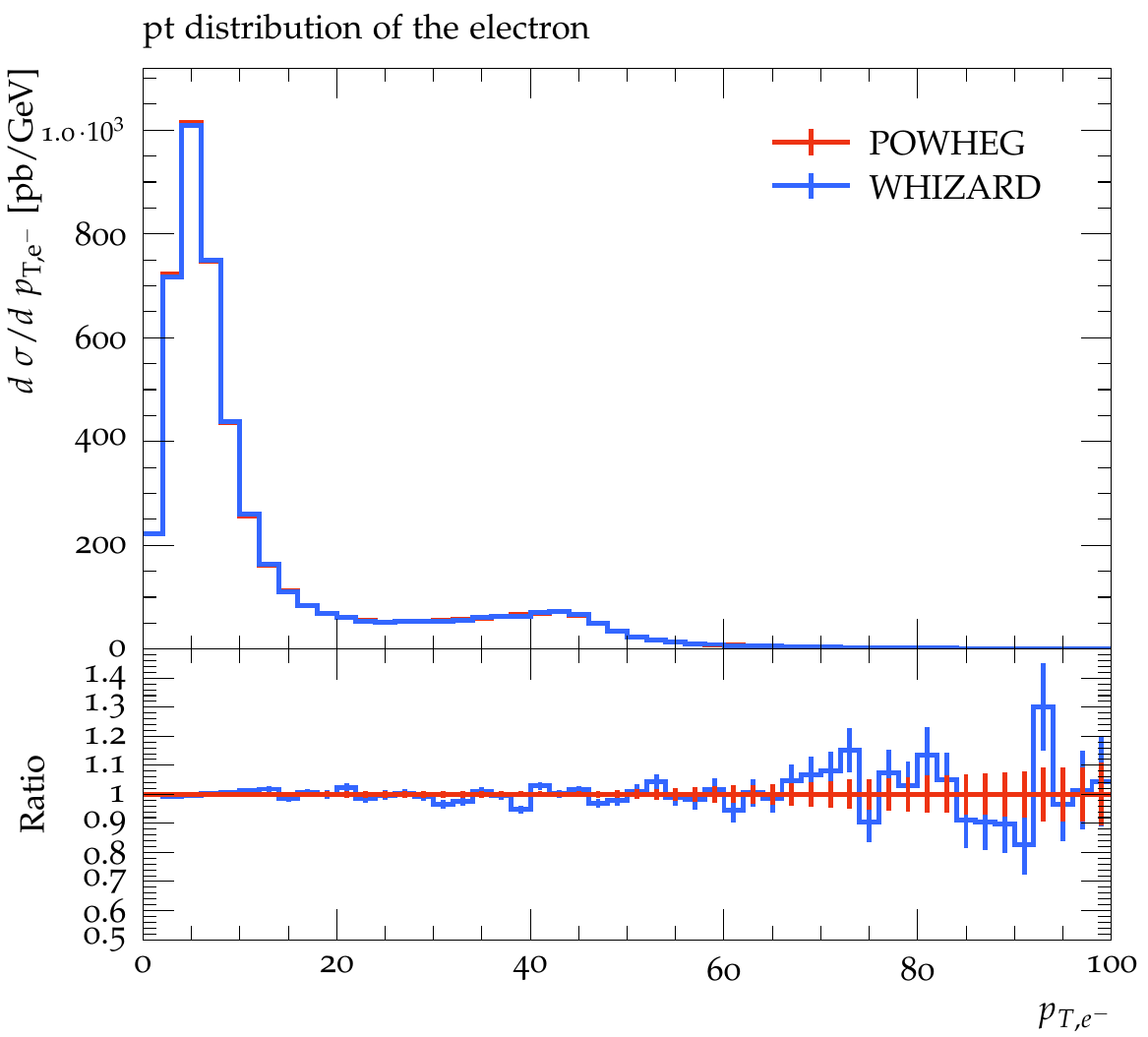}
	\includegraphics[width=0.3\textwidth]{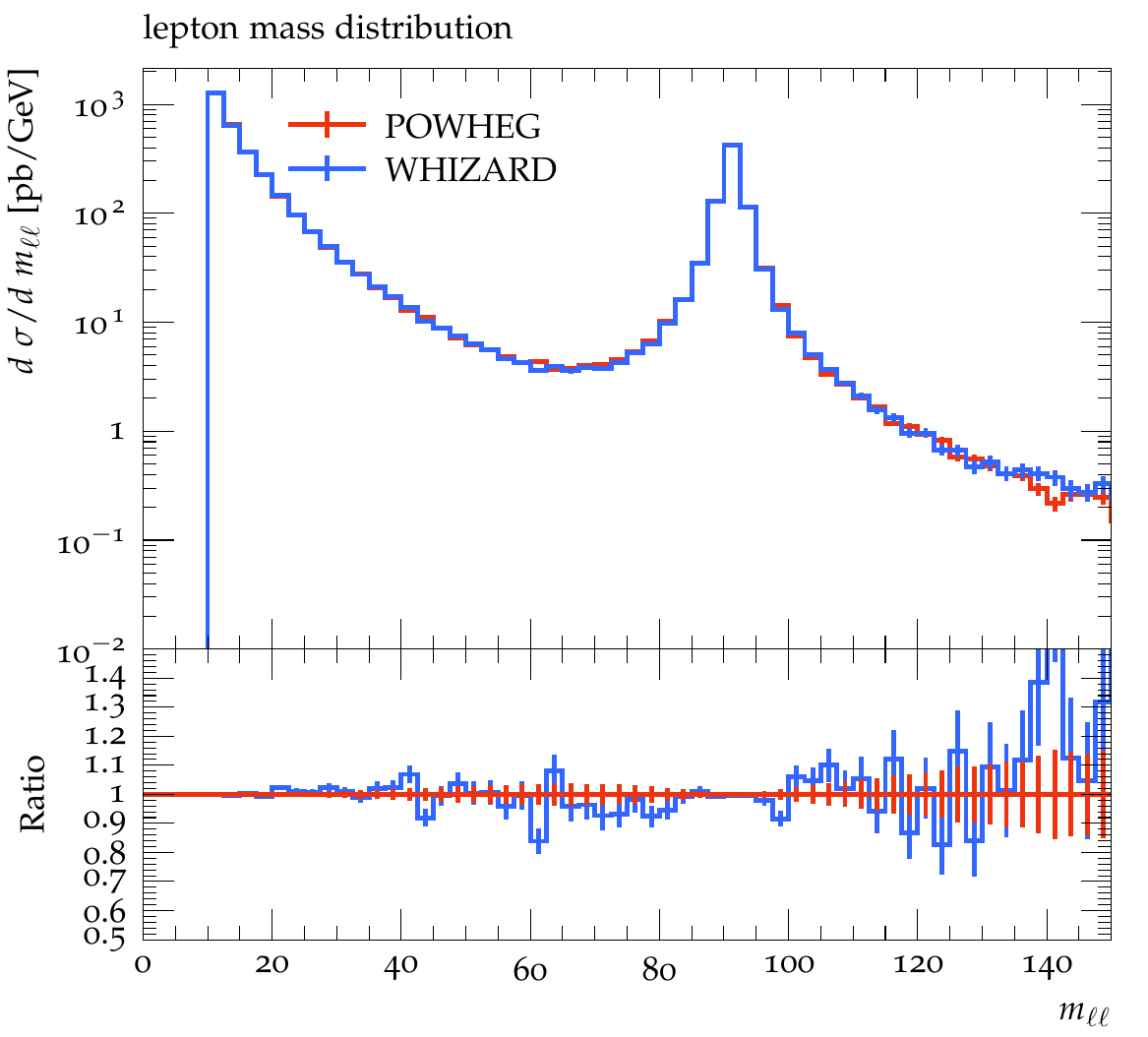}
	\includegraphics[width=0.3\textwidth]{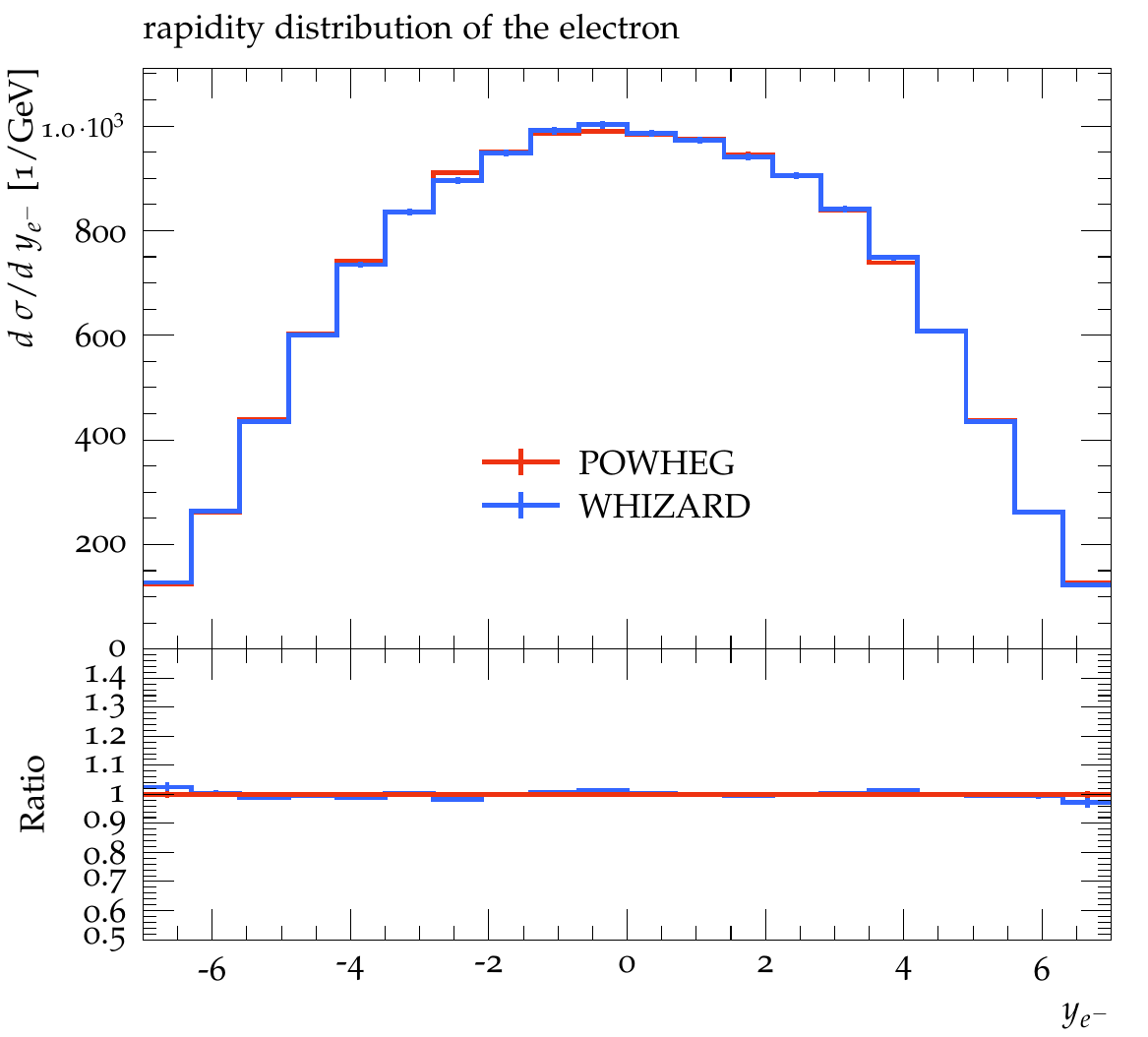}
	\caption{Distributions at NLO QCD to $pp\rightarrow e^+e^-$ for the validation of matched and showered events}
	\label{Powheg}
\end{figure}

The work on the POWHEG matching in \texttt{WHIZARD} started with earlier studies on $e^+e^-\rightarrow t\bar{t}(H)$ precision calculations \cite{ChokoufeNejad:2015kpc}. Aiming for a generalization of the
POWHEG matching to arbitrary processes the implementation now has been extended to $pp$ processes with validation completed for Drell-Yan and similar processes. Comparisons of $p_{T,e^-}$, $m_{e^+e^-}$ and $y_{e^-}$ distributions for $pp\rightarrow e^+e^-$ with POWHEG-matched events from \texttt{WHIZARD} and \texttt{POWHEG-BOX} \cite{Alioli:2008gx} and showered with \texttt{PYTHIA} \cite{Sjostrand:2014zea} are shown in Fig.~\ref{Powheg}.
\section{Outlook}
Future plans include the application of NLO-NLL electron PDFs in lepton collision observables at NLO as well as efficiency improvements on the phase-space generation for NLO EW calculations to high multiplicity processes at the LHC. Furthermore, the automated POWHEG matching is supposed to be extended accounting for NLO corrections in the full SM.
\section*{Acknowledgements}
This work was funded by the Deutsche Forschungsgemeinschaft under Germany’s Excellence Strategy
– EXC 2121 ”Quantum Universe” - 390833306. For cross-checks with \texttt{MUNICH} and numerous advices on automated NLO EW corrections PB wants to thank Stefan Kallweit.


\begin{thebibliography}
	\footnotesize
\setlength{\itemsep}{1pt}
\bibitem{Kilian:2007gr}
W.~Kilian, \emph{et. al.} 
Eur. Phys. J. C \textbf{71} (2011), 1742
arXiv: \href{https://arxiv.org/abs/0708.4233}{0708.4233~[hep-ph]}.
\bibitem{Moretti:2001zz}
M.~Moretti, \emph{et. al.} 
arXiv: \href{https://arxiv.org/abs/hep-ph/0102195}{hep-ph/0102195~[hep-ph]}.
\bibitem{Brass:2018xbv}
S.~Brass, \emph{et. al.} 
Eur. Phys. J. C \textbf{79} (2019) no.4, 344
arXiv: \href{https://arxiv.org/abs/1811.09711}{1811.09711~[hep-ph]}.
\bibitem{Buccioni:2019sur}
F.~Buccioni, \emph{et. al.} 
Eur. Phys. J. C \textbf{79} (2019) no.10, 866
arXiv: \href{https://arxiv.org/abs/1907.13071}{1907.13071~[hep-ph]}.
\bibitem{Actis:2012qn}
S.~Actis, \emph{et. al.} 
JHEP \textbf{04} (2013), 037
arXiv: \href{https://arxiv.org/abs/1211.6316}{1211.6316~[hep-ph]}.
\bibitem{Cullen:2014yla}
G.~Cullen, \emph{et al.}
Eur. Phys. J. C \textbf{74} (2014) no.8, 3001
arXiv: \href{https://arxiv.org/abs/1404.7096}{1404.7096~[hep-ph]}.
\bibitem{Bach:2017ggt}
F.~Bach, \emph{et. al.} 
JHEP \textbf{03} (2018), 184
arXiv: \href{https://arxiv.org/abs/1712.02220}{1712.02220 [hep-ph]}.
\bibitem{ChokoufeNejad:2016qux}
B.~Chokouf\'e Nejad, \emph{et. al.}
JHEP \textbf{12} (2016), 075
[arXiv:1609.03390 [hep-ph]].
\bibitem{Frederix:2018nkq}
R.~Frederix, \emph{et. al.} 
JHEP \textbf{07} (2018), 185
arXiv: \href{https://arxiv.org/abs/1804.10017}{1804.10017~[hep-ph]}.
\bibitem{Kallweit:2014xda}
S.~Kallweit, \emph{et. al.} 
JHEP \textbf{04} (2015), 012
arXiv: \href{https://arxiv.org/abs/1412.5157}{1412.5157~[hep-ph]}.
\bibitem{Buonocore:2021rxx}
L.~Buonocore, \emph{et. al.} 
Phys. Rev. D \textbf{103} (2021), 114012
arXiv: \href{https://arxiv.org/abs/2102.12539}{2102.12539~[hep-ph]}.

\bibitem{Bredt:2022dmm}
P.~Bredt, \emph{et. al.} 
arXiv: \href{https://arxiv.org/abs/2208.09438}{2208.09438~[hep-ph]}.
\bibitem{Frixione:2019lga}
S.~Frixione,
JHEP \textbf{11} (2019), 158
arXiv:
\href{https://arxiv.org/abs/1909.03886}{1909.03886 [hep-ph]}.
\bibitem{Bertone:2019hks}
V.~Bertone, \emph{et. al.} 
JHEP \textbf{03} (2020), 135
arXiv:
\href{https://arxiv.org/abs/1911.12040}{1911.12040 [hep-ph]}.
\bibitem{ChokoufeNejad:2015kpc}
B.~Chokoufe Nejad, \emph{et. al.} 
PoS \textbf{EPS-HEP2015} (2015), 317
arXiv:
\href{https://arxiv.org/abs/1510.02739}{1510.02739 [hep-ph]}.
\bibitem{Alioli:2008gx}
S.~Alioli, \emph{et. al.} 
JHEP \textbf{07} (2008), 060
arXiv: \href{https://arxiv.org/abs/0805.4802}{0805.4802 [hep-ph]}.
\bibitem{Sjostrand:2014zea}
T.~Sj\"ostrand, \emph{et. al.} 
Comput. Phys. Commun. \textbf{191} (2015), 159-177
arXiv: \href{https://arxiv.org/abs/1410.3012}{1410.3012~[hep-ph]}.
\end{thebibliography}
\end{document}